\documentclass[aps, prl, superscriptaddress, reprint]{revtex4-1}

\usepackage{graphicx} 
\usepackage{bm}
\usepackage{natbib}
\usepackage{verbatim}
\usepackage{epstopdf}
\usepackage{hyperref}

\begin{document}
\title{LO-phonon emission rate of hot electrons from an on-demand single-electron source in a GaAs/AlGaAs heterostructure}

\author  {N. Johnson}
\email[]{nathan.johnson@lab.ntt.co.jp}
\altaffiliation[Present address: ]{NTT Basic Research Laboratories, NTT Corporation, 3-1 Morinosato Wakamiya, Atsugi, Kanagawa 243-0198, Japan}
\affiliation{National Physical Laboratory, Hampton Road, Teddington, Middlesex TW11 0LW, United Kingdom}
\affiliation{London Centre for Nanotechnology, and Department of Electronic and Electrical Engineering, University College London, Torrington Place, London, WC1E 7JE, United Kingdom}

\author {C. Emary}
\affiliation{Joint Quantum Centre Durham-Newcastle, School of Mathematics and Statistics, Newcastle University, Newcastle upon Tyne NE1 7RU, United Kingdom.}
\author{S. Ryu}
\affiliation{Department of Physics, Korea Advanced Institute of Science and Technology, Daejeon 305-701, Republic of Korea}
\author{H.-S. Sim}
\affiliation{Department of Physics, Korea Advanced Institute of Science and Technology, Daejeon 305-701, Republic of Korea}
\author {P. See}
\affiliation{National Physical Laboratory, Hampton Road, Teddington, Middlesex TW11 0LW, United Kingdom}
\author {J.D. Fletcher}
\affiliation{National Physical Laboratory, Hampton Road, Teddington, Middlesex TW11 0LW, United Kingdom}
\author{J.P. Griffiths}
\affiliation {Cavendish Laboratory, University of Cambridge, J.J. Thomson Avenue, Cambridge CB3 0HE, United Kingdom}
\author{G.A.C. Jones}
\affiliation{Cavendish Laboratory, University of Cambridge, J.J. Thomson Avenue, Cambridge CB3 0HE, United Kingdom}
\author{I. Farrer}
\affiliation{Department of Electronic and Electrical Engineering, University of Sheffield, Mappin Street, Sheffield S1 3JD, United Kingdom}
\author{D.A. Ritchie}
\affiliation{Cavendish Laboratory, University of Cambridge, J.J. Thomson Avenue, Cambridge CB3 0HE, United Kingdom}
\author{M. Pepper}
\affiliation{London Centre for Nanotechnology, and Department of Electronic and Electrical Engineering, University College London, Torrington Place, London, WC1E 7JE, United Kingdom}
\author{T.J.B.M Janssen}
\affiliation{National Physical Laboratory, Hampton Road, Teddington, Middlesex TW11 0LW, United Kingdom}
\author{M. Kataoka}
\email[]{masaya.kataoka@npl.co.uk}
\affiliation{National Physical Laboratory, Hampton Road, Teddington, Middlesex TW11 0LW, United Kingdom}

\begin{abstract}
Using a recently-developed time-of-flight measurement technique with 1~ps time resolution and electron-energy spectroscopy, we developed a method to measure the longitudinal-optical-phonon emission rate of hot electrons travelling along a depleted edge of a quantum Hall bar. 
A comparison of the experimental results to a single-particle model implies that the main scattering mechanism involves a two-step process via intra-Landau-level transition.
We show this scattering can be suppressed by controlling the edge potential profile, and a scattering length $>$~1 mm can be achieved, allowing the use of this system for scalable single-electron device applications.
\end{abstract}



\maketitle

The development of accurate, on-demand, hot single electron sources has opened up a new energy domain in which to study fundamental electron behaviour in the solid state \cite{Blumenthal, Kaestner, Leicht, Giblin}. 
In particular, this energy domain presents a unique environment in which we can study the nature of single-particle physics.
One extensively studied phenomenon of hot electrons in GaAs systems is the energy relaxation by emission of longitudinal optical (LO) mode phonons \cite{Heiblum, dasSarma, dasSarma2, Sivan, Taubert, Taubert2}.
This causes the electrons to undergo a discrete energy loss ($\hbar \omega_{LO} \sim$~36~meV) per emission of one LO phonon.
This inelastic scattering process is one of the reasons that high-energy quasiparticles are considered to be inappropriate for experiments that require electron coherence, such as interferometry.
Suppression of this scattering mechanism creates a possibility to utilise these devices for experiments such as electron quantum optics \cite{Bocquillon} and technological applications \cite{Bennett}.
Therefore, it is important to characterise the rate of LO-phonon scattering and understand its mechanisms.
However, it has so far not been straightforward to measure the rate directly in transport measurements, because although it is relatively easy to measure the scattering probability, it is not easy to measure the electron velocity.
Moreover, in the presence of a background Fermi sea, electron-electron interactions strongly affect the system, masking the simple single-particle physics \cite{Taubert2}.
Now with the development of on-demand hot single-electron sources \cite{Blumenthal} and electron energy spectroscopy \cite{Jon}, it is possible to perform energy- and time-resolved study of hot electron transport through an intrinsic/depleted region free from electron-electron interactions.

In this work we present measurements of the LO-phonon emission rate of hot electrons travelling in quantum Hall edge states.
In order to deduce the LO phonon emission rate, we first measure the LO phonon emission probability $P_{LO}^{l}$ along an electron path with length $l$ using energy spectroscopy \cite{Jon}. 
Then, we measure the average electron drift velocity $v_{d}$ (or electron travel time $\tau = l/v_{d}$) in the same path using a time-of-flight measurement \cite{masaya}.
The scattering rate $\Gamma_{LO}$ is calculated as $\Gamma_{LO} =  - \frac{v_{d}}{l} \rm{ln}$$(1-P_{LO}^{l})$.
We perform a detailed study of how the emission rate varies as electron energy, edge-potential profile and magnetic field are varied.
Comparison with theory for direct LO-phonon emission \cite{Emary} shows that, while qualitatively in agreement, our measured rate is many orders of magnitude larger than the predicted values. 
We suggest that an enhanced emission process due to an inter-Landau-level transition may be present \cite{Komiyama}.
Moreover, we demonstrate that the scattering length can be made as large as 1~mm by controlling the edge potential.

\begin{figure*}[ht!]
	\includegraphics[width=\linewidth]{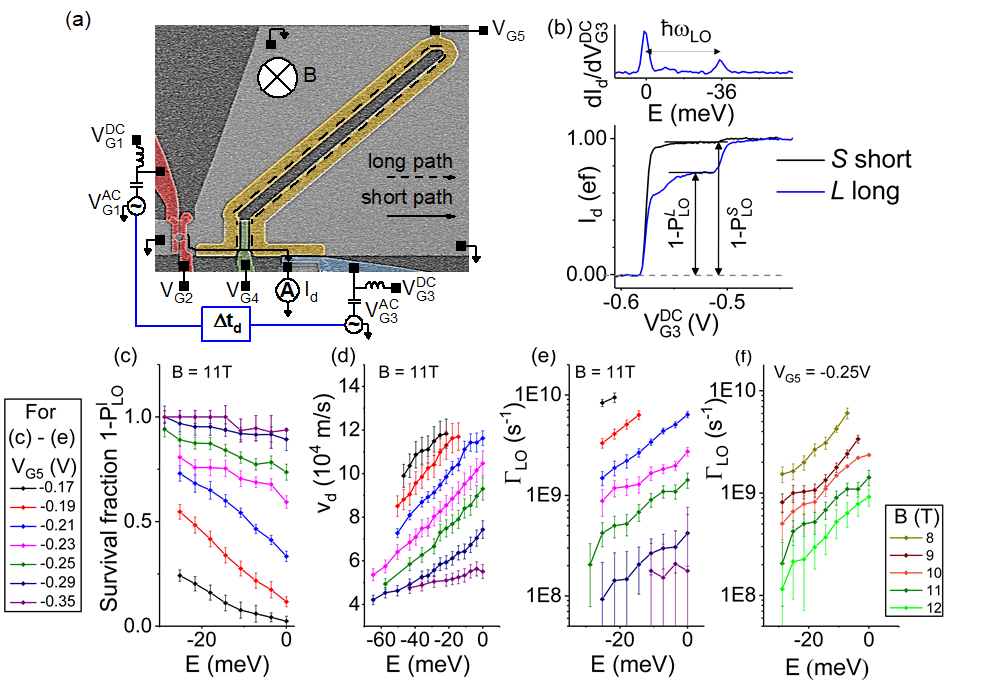}
	\caption{
		(a) SEM image of the device and electrical connections. The 2DEG region is shaded in light grey. Metallic gates are coloured - electron pump $\rm{G_{1}}$, $\rm{G_{2}}$, red; detector gate $\rm{G_{3}}$, blue; depletion gate $\rm{G_{5}}$, yellow; deflection gate $\rm{G_{4}}$, green. 
		Note that the area encircled by the yellow gate, $\rm{G_{5}}$, is etched away. 
		Long and short paths are marked with dashed (solid) lines respectively.
		(b) Measurement of phonon emission probability as a fraction of pumped current, for the case of the current travelling short path S (black) and long path L (blue).
		The survival fraction $1-P_{0}^{L(S)}$ is defined as the fraction of the detector current ($I_{d}$) at the phonon plateau against $ef$.
		Above, the derivative $dI_{d}/dV_{G3}^{DC}$ of the long path trace shows the energy spectrum of electrons with original emission energy (left peak) and the ones that have emitted one LO phonon (right peak), which are separated by the LO-phonon energy of 36~meV.
		(c) Survival fraction $1-P_{LO}^{l}$ as a function of electron emission energy $E$, for different values of $V_{G5}$. 
		(d) Electron drift velocity $v_{d}$ as a function of $E$, for different values of $V_{G5}$. 
		(e) LO-phonon emission rate, $\Gamma_{LO} = -\frac{v_{d}}{l} \rm{ln}$$(1-P_{0}^{l})$, as a function of $E$, for different values of $V_{G5}$. 
		(f) LO-phonon emission rate as a function of $E$, measured at different magnetic fields $B$ and at $V_{G5} = -0.25$~V.
	}
	\label{fig:first}
\end{figure*}

A scanning-electron-microscope (SEM) image of an identical device to that used, with schematic electrical connections, is shown in Fig.~\ref{fig:first}(a).
The sample has a similar geometry to those used in Ref.~\cite{masaya}.
A two-dimensional electron gas (2DEG) is defined 90~nm below the surface of a GaAs/AlGaAs heterostructure.
Parts of the substrate are chemically etched using electron-beam lithography to define the mesa shown in light grey in Fig.~\ref{fig:first}(a) 
(including the area encircled by the yellow gate).
Five Au/Ti metallic gates are patterned onto the surface using electron-beam lithography.
Gates G$_{\rm{1}}$ and G$_{\rm{2}}$, shaded in red, define the quantum dot single-electron pump that acts as the source of energy-tunable hot electrons \cite{Blumenthal, Kaestner, Leicht, Giblin}.
G$_{\rm{1}}$ is driven by an ac sinusoidal waveform $V_{G1}^{AC}$ at frequency $f = 400$~MHz with a peak-to-peak amplitude $\sim$1~V from one channel of an arbitrary waveform generator (AWG), in addition to a dc voltage $V_{G1}^{DC}$.
We tune the gate voltages to pump one electron per cycle, producing a current $I_{p} = ef \approx 64$~pA, with $e$ the elementary charge.

In the presence of a perpendicular magnetic field $B$, the electrons emitted from the pump travel along the sample edge either in the short (5~$\mu$m) or long path (28~$\mu$m) indicated by the solid and dashed lines respectively in Fig.~\ref{fig:first}(a), just as in the edge-state transport in the quantum Hall regime \cite{Halperin}, but with an energy (about 100~meV) above the Fermi energy $E_{F}$ \cite{Jon}.
The voltage $V_{G4}$ applied to the deflection gate $\rm{G_{4}}$ determines which path the electrons take 
\footnote{This is the lithographic length, and we assume a $\pm 5\%$ error because we do not know the exact path the electrons take. We note that the error in path length estimate does not contribute to the error in the measurements of phonon emission rate as it is cancelled out by both the velocity and length estimates. It does, however, contribute to the error in the estimate of the edge potential profile.}. 
The time-of-flight method in Ref.~\cite{masaya} gives the electron velocity in the loop section (the part that encircles the elongated etched area and does not include the paths along the gate $\rm{G_{4}}$) of length $l \approx 20$~$\mu$m (see supplemental material for determination of this length). 
This is the length where we investigate the LO-phonon emission rate. 

Gate $\rm{G_{3}}$ (blue) is the detector gate. 
A dc gate voltage $V_{G3}^{DC}$ is used to detect the electron energy \cite{Jon} and the LO phonon emission probability by measuring the transmitted detector current $I_{d}$.
For the time-of-flight measurements, an ac square waveform $V_{G3}^{AC}$, with controllable time delay $\Delta t_{d}$, is applied in addition to $V_{G3}^{DC}$ \cite{masaya, joanna, Johnson, Kataoka}.
Gate G$_{\rm{5}}$ (yellow) is the depletion gate, and is used to change the potential profile at the edge where electrons propagate by applying a dc voltage $V_{G5}$.
We note that, throughout this work, the voltage applied to the depletion gate is negative enough to deplete the 2DEG underneath, but not negative enough to push the path of our hot electrons outside the gated region. 
Experiments are performed in a dilution refrigerator with a base temperature $\sim$~30~mK and with a perpendicular magnetic field of 8~-~12~T. 

\begin{figure}
	\includegraphics[width=\linewidth]{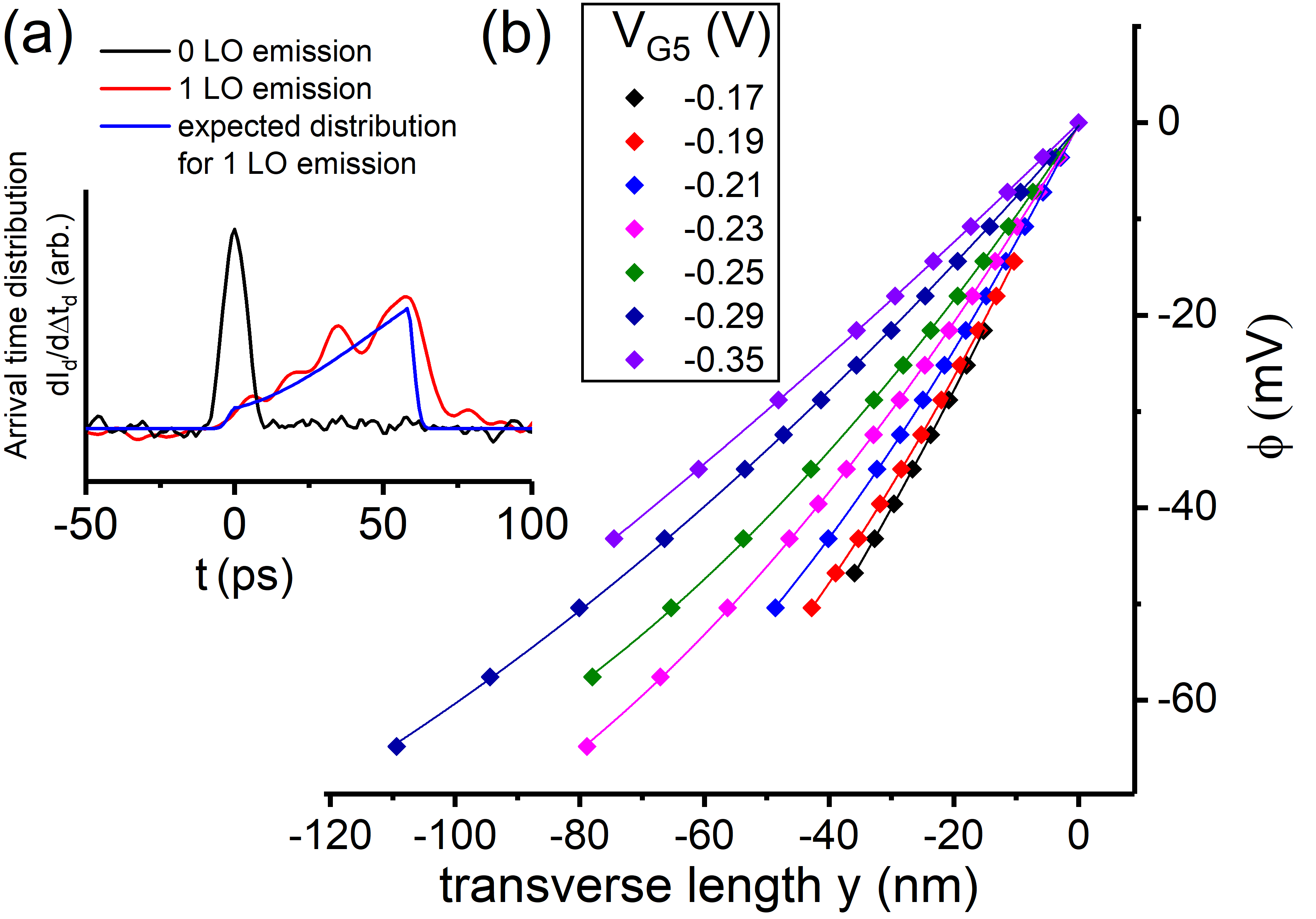}
	\caption{
	(a) The arrival time distributions $A_{LO} (t)$ for electrons emitting 0 (black) and 1 (red) LO-phonons.
	The good match to the expected distribution (blue) implies that the phonon emission rate is uniform.
	(b) The edge potential profile $\phi$, measured for different $V_{G5}$.
	We take $(y,\phi) = (0,0)$ as the point of highest electron emission energy attainable by the pump.
	Solid lines show a parabolic fit, which is used to calculate the phonon emission rate.
	}
	\label{fig:second}
\end{figure}

In order to determine the LO-phonon emission rate, we first measure the probability of emission in the long (short) path $P_{LO}^{L(S)}$.
This is done by measuring $I_{d}$ while sweeping the detector $V_{G3}^{DC}$.
This gives us a certain fraction of pumped current $I_{p}$, which is the proportion of pumped electrons that retain enough energy to pass the detector barrier.
Fig.~\ref{fig:first}(b) shows an example of this measurement taken for the short (black) and long (blue) paths
\footnote{In this example, $V_{G2} = -0.44$~V, $V_{G5} = -0.25$~V, $B = 11$~T, and $V_{G4} = -0.3$~V (short path, $S$) or $-0.65$~V (long path, $L$).}.
As $V_{G3}^{DC}$ is swept to a more positive value (i.e. as the detector barrier is lowered), $I_{d}$ increases and shows a sub-step (the phonon plateau) before rising to $ef$ and saturating. 
The derivative $dI_{d}/dV_{G3}^{DC}$ shows two peaks as shown in the top panel in Fig.~\ref{fig:first}(b). 
From Ref~\cite{Jon}, we identify the left-hand peak as the energy spectrum of electrons at the original emission energy, and the  right-hand peak as that of electrons that have emitted one LO phonon before arriving at the detector. 
From this, we can convert the $V_{G3}^{DC}$ scale into the energy scale, assuming $\hbar \omega_{LO} \sim 36$~meV. 
The height of the phonon plateau (when normalised to $ef$) gives a survival fraction of electrons arriving at the detector without any LO-phonon emission, i.e. $1 - P_{LO}^{L(S)}$ for long (short) path.

We then deduce the survival fraction in the loop alone by noting that the long path survival fraction is the product of the survival fractions of the short path and the loop, i.e. $1-P_{LO}^{l} = (1-P_{LO}^{L}) /(1- P_{LO}^{S})$.
With our single-electron source, the electron emission energy can be tuned by changing the height of the exit barrier, i.e. by varying $V_{G2}$ \cite{Jon}.
Fig.~\ref{fig:first}(c) shows $1 - P_{LO}^{l}$ as a function of the electron emission energy $E$ (here, we define $E=0$ for the highest emission energy used) taken with a different value of  depletion gate voltage $V_{G5}$, at $B= 11$~T. 
We also measure the average electron drift velocity $v_{d}$ (of the non-LO-phonon emitting fraction) in the loop section for different electron emission energy using the method described in Ref.~\cite{masaya} [Fig.~\ref{fig:first}(d)].
(We will later calculate the form of the edge potential profile from these data.)
Assuming the constant rate of LO phonon emission $\Gamma_{LO}$ throughout the loop, we can calculate the rate as $\Gamma_{LO} = - \frac{v_{d}}{l} \rm{ln}$$(1-P_{LO}^{l})$  [Fig.~\ref{fig:first}(e)]
\footnote{We note that there is a smaller number of data points in Fig.~\ref{fig:first}(e) compared to Fig.~\ref{fig:first}(c) and (d). This is because we need both $1-P_{LO}^{l}$ and $v_{d}$ available in order to deduce $\Gamma_{LO}$ for a given experimental condition. The range in $E$ for which $1-P_{LO}^{l}$ can be measured is limited because, as the original electron emission is lowered below $E = -25$~meV (the energy of the electron that emits one phonon will be below -61~meV), we cannot reliably measure the height of the sub-step in $I_{d}$ as a spurious current due to a pick up of RF signal by the 2DEG starts to flow through the detector barrier when it is made too low. Also, $v_{d}$ measurement is difficult for less negative $V_{G5}$ and higher $E$, where the LO phonon emission rate is high, as there are not enough electrons reaching the detector with the original emission energy, so cannot be detected by the measurement of $I_{d}$.}.
We observe that the rate of phonon emission is strongly a function of both emission energy $E$ and $V_{G5}$.
In Fig.~\ref{fig:first}(f) we repeat this analysis at various $B$ values, for the case $V_{G5} = -0.25V$.
We also clearly see a strong field dependence on the rate of phonon emission, with emission suppressed at higher fields \cite{Jon}.
We note that the reason that there is no data below $\Gamma_{LO} = 10^{8} s^{-1}$ in Fig.~\ref{fig:first}(e) is not because we cannot tune the device into that regime, but because the scattering probability becomes so small that it is difficult to detect it by the measurement of the detector current. 
If we extrapolate the data for $V_{G5}=-0.35$~V to $E=-20$~meV [experimental data for survival probability in this configuration is shown in Fig.~\ref{fig:first}(c)], $\Gamma_{LO} \sim 5 \times 10^7$~$\rm{s}^{-1}$ and $v_{d} \sim 5 \times 10^4$~m/s. 
Hence, we expect the scattering length $v_{d}/\Gamma_{LO}$ to be $\sim 1$~mm in this configuration. 
The scattering length can easily be made even longer.

Because we have temporal detection of the electrons, we can detect whether the phonon emission happens uniformly around the ring, or whether there are phonon emission `hot spots'. 
If the latter is the case, the measured phonon emission rate by our technique may not represent the actual emission rate. 
A phonon-emission hot spot will produce a peak in the arrival-time distribution, $A_{LO}(t) \propto dI_{d}/d\Delta t_{d}$ of the electrons that have emitted one LO phonon. 
This peak in $A_{LO}(t)$ is expected to occur later than the peak of the original electrons, as the electrons at a lower energy are expected to travel at a lower velocity.
However, if the phonon emission rate is constant throughout the path, we expect the arrival-time distribution to spread out, as electrons emit phonons at various parts of the path. 
In Fig.~\ref{fig:second}(a), the measurement of arrival time distribution, using the method given in Refs. \cite{joanna, Kataoka}, is shown for the electrons with original emission energy (black) and the electrons that have emitted one phonon (red).  
The expected form of $A_{LO}(t)$ (blue) with constant emission rate as calculated from Ref.~\cite{Emary} (and see supplemental material), and shows a good match with the experimental data.
This implies that our assumption of constant phonon emission rate is reasonable.  

Now we investigate if our experimental results can be explained within the framework of an existing single-phonon emission model \cite{dasSarma, dasSarma2, Emary}.
To compare the measured rate against theory, we need to know the shape of the potential profile $\phi$, from which we can calculate the electronic wave function.
A key parameter in the calculation of emission rates is the edge potential profile, as this dictates the spatial position of the electron before and after emission.
This can be deduced from the drift velocity measurements using the method described in Ref.~\cite{masaya}, and we plot the deduced potential profile in Fig.~\ref{fig:second}(b), for each case of $V_{G5}$ we have measured the LO-phonon rate, under the same conditions as the emission rate was measured.

\begin{figure}
	\includegraphics[width=\linewidth]{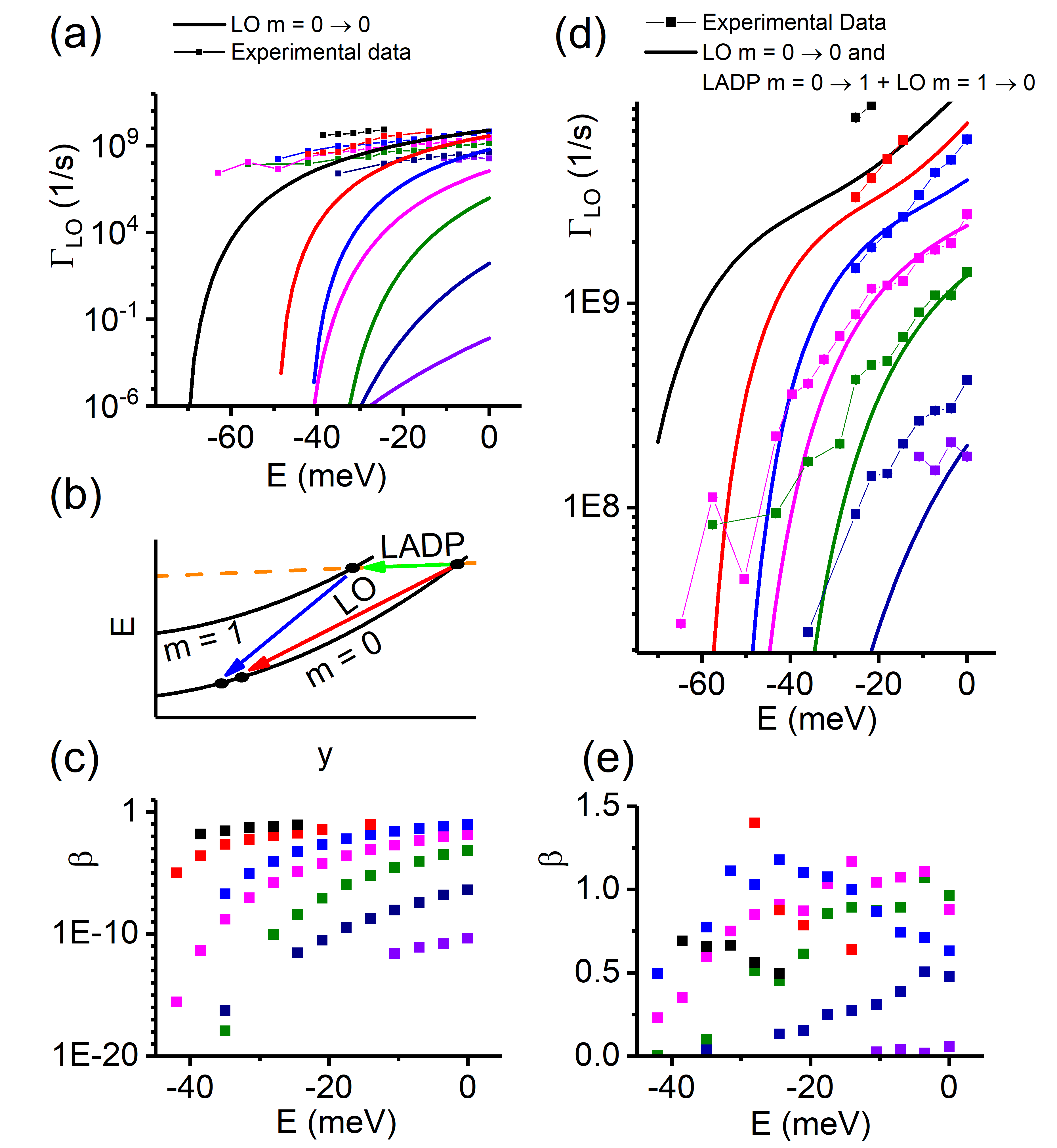}
	\caption{
		(a) Comparison of measured LO-phonon emission rates (squares, colours match Fig.~\ref{fig:first}(c)) with the calculated rates for $m = 0 \rightarrow 0$ LO-phonon mode (solid lines) show poor agreement.
		(b) Two lowest Landau levels at the sample edge ($m=0$ and $1$), showing the phonon scattering paths. Red arrow: $m=0 \rightarrow m=0$ LO-phonon emission, green arrow: $m=0 \rightarrow 1$ LADP phonon emission, and blue arrow: $m=1 \rightarrow 0$ LO phonon emission.
		(c) The discrepancy fraction $\beta$ = expected rate/measured rate of phonon emission for the direct ($m=0\to0$) mode.
		(d) As (a), with the combined rate of emission from the $m=0 \rightarrow 0$ LO-mode and the $m=0 \rightarrow 1$ LADP emission then $m =1 \rightarrow 0$ LO emission mode (solid line), shows better agreement with the data.
		(e) The discrepancy fraction $\beta$ for the combined mode of direct and the two step process. In this combined case, we see that we are much closer to unity. 
	}
	\label{fig:third}
\end{figure}

The theoretical LO-phonon emission rate can be calculated using the method described in Ref.~\cite{Emary}, by performing the harmonic fit to the experimentally determined potential profiles, $\phi$. 
[We note that we use a z-confinement (in the substrate growth direction) of 10~nm, the variation of which within 50\% does not change the order of magnitude of the result, and is the only free parameter in the model.]
Fig.~\ref{fig:third}(a) plots the experimental data [square symbols, same as that presented in Fig.~\ref{fig:first}(e)] and calculated LO phonon emission rate (solid lines) for intra-Landau-level transition ($m=0 \rightarrow 0$, where $m$ is the Landau level index) [red arrow in Fig.~\ref{fig:third}(b)] for different values of $V_{G5}$.
In Fig.~\ref{fig:third}(c), the ratio $\beta$ of the theoretically calculated value and experimental value is plotted, showing a large discrepancy by many orders of magnitude.
When only direct LO-phonon emission is considered, the predicted rates of emission are far lower than those measured experimentally.
In order to explain this discrepancy, we consider the possibility that there are other paths to phonon emission.

One such possibility is a process in which an electron is first transferred to the $m=1$ Landau level via longitudinal acoustic via deformation potential (LADP) mode \cite{Climente, Komiyama} and then is transferred back to the $m=0$ Landau level via LO-phonon emission [see the green and blue arrows in Fig.~\ref{fig:third}(b)]. 
This two-step process can be faster than the direct $m = 0\to0$ transition. 
This is because the rate of any particular transition is proportional to the electronic wave function overlap and thus to a factor $e^{-(\Delta y/l_{b})^2/2}$, where $\Delta y $ is the change in the guide-centre coordinate in the transition and $l_{b}$ is the magnetic length \cite{Emary}.  
Since $\Delta y$ can be smaller in each of the emissions in the two-stage process, an exponential speed-up is gained and in some circumstances, this can overcome the inherent slowness of a two step process.
The sum of the direct and two-step rates are shown in Fig.~\ref{fig:third}(d), and the ratio $\beta$ is plotted in Fig.~\ref{fig:third}(e). 
This clearly shows that the calculated curves agree with experiments within an order of magnitude, except for those with the most negative values of $V_{G5}$.
Here, the edge-confinement potential is shallower, and harmonic approximation may be inadequate, which may be the origin of the discrepancy.

In summary, we have demonstrated a detailed study of LO-phonon emission rate of hot electrons in quantum Hall edge states.
We measured this rate at different electron energies, magnetic fields, and under different potential profiles, which is controlled by an edge depletion gate that covers the whole electron path from pump to detector.
The depletion of the background electron gas allows us to study these effects within a simple single-particle picture.
Comparisons with theory suggest that inter-Landau-level scattering via acoustic phonon emission is involved in the LO-phonon emission. 
We found that the phonon emission rate can be controlled by the edge-depletion gate, and scattering length can be enhanced as much as 1~mm or even longer, which enables the construction of large-scale single-electron devices.

We thank S. P. Giblin  and S. Ludwig for useful discussions.
This research was supported by the UK Department for Business, Energy, and Industrial Strategy, the UK EPSRC, and by the project EMPIR 15SIB08 e-SI-Amp. 
This project has received funding from the EMPIR programme co-financed by the Participating States and from the European Union’s Horizon 2020 research and innovation programme and by Korea NRF (Grant No. 2016R1A5A1008184).

\bibliography{phonons_reduced}	

\begin{thebibliography}{24}%
\makeatletter
\providecommand \@ifxundefined [1]{%
 \@ifx{#1\undefined}
}%
\providecommand \@ifnum [1]{%
 \ifnum #1\expandafter \@firstoftwo
 \else \expandafter \@secondoftwo
 \fi
}%
\providecommand \@ifx [1]{%
 \ifx #1\expandafter \@firstoftwo
 \else \expandafter \@secondoftwo
 \fi
}%
\providecommand \natexlab [1]{#1}%
\providecommand \enquote  [1]{``#1''}%
\providecommand \bibnamefont  [1]{#1}%
\providecommand \bibfnamefont [1]{#1}%
\providecommand \citenamefont [1]{#1}%
\providecommand \href@noop [0]{\@secondoftwo}%
\providecommand \href [0]{\begingroup \@sanitize@url \@href}%
\providecommand \@href[1]{\@@startlink{#1}\@@href}%
\providecommand \@@href[1]{\endgroup#1\@@endlink}%
\providecommand \@sanitize@url [0]{\catcode `\\12\catcode `\$12\catcode
  `\&12\catcode `\#12\catcode `\^12\catcode `\_12\catcode `\%12\relax}%
\providecommand \@@startlink[1]{}%
\providecommand \@@endlink[0]{}%
\providecommand \url  [0]{\begingroup\@sanitize@url \@url }%
\providecommand \@url [1]{\endgroup\@href {#1}{\urlprefix }}%
\providecommand \urlprefix  [0]{URL }%
\providecommand \Eprint [0]{\href }%
\providecommand \doibase [0]{http://dx.doi.org/}%
\providecommand \selectlanguage [0]{\@gobble}%
\providecommand \bibinfo  [0]{\@secondoftwo}%
\providecommand \bibfield  [0]{\@secondoftwo}%
\providecommand \translation [1]{[#1]}%
\providecommand \BibitemOpen [0]{}%
\providecommand \bibitemStop [0]{}%
\providecommand \bibitemNoStop [0]{.\EOS\space}%
\providecommand \EOS [0]{\spacefactor3000\relax}%
\providecommand \BibitemShut  [1]{\csname bibitem#1\endcsname}%
\let\auto@bib@innerbib\@empty
\bibitem [{\citenamefont {Blumenthal}\ \emph {et~al.}(2007)\citenamefont
  {Blumenthal}, \citenamefont {Kaestner}, \citenamefont {Li}, \citenamefont
  {Giblin}, \citenamefont {Janssen}, \citenamefont {Pepper}, \citenamefont
  {Anderson}, \citenamefont {Jones},\ and\ \citenamefont
  {Ritchie}}]{Blumenthal}%
  \BibitemOpen
  \bibfield  {author} {\bibinfo {author} {\bibfnamefont {M.~D.}\ \bibnamefont
  {Blumenthal}}, \bibinfo {author} {\bibfnamefont {B.}~\bibnamefont
  {Kaestner}}, \bibinfo {author} {\bibfnamefont {L.}~\bibnamefont {Li}},
  \bibinfo {author} {\bibfnamefont {S.}~\bibnamefont {Giblin}}, \bibinfo
  {author} {\bibfnamefont {T.~J. B.~M.}\ \bibnamefont {Janssen}}, \bibinfo
  {author} {\bibfnamefont {M.}~\bibnamefont {Pepper}}, \bibinfo {author}
  {\bibfnamefont {D.}~\bibnamefont {Anderson}}, \bibinfo {author}
  {\bibfnamefont {G.}~\bibnamefont {Jones}}, \ and\ \bibinfo {author}
  {\bibfnamefont {D.~A.}\ \bibnamefont {Ritchie}},\ }\href@noop {} {\bibfield
  {journal} {\bibinfo  {journal} {Nature Phys. 3 343}\ } (\bibinfo {year}
  {2007})}\BibitemShut {NoStop}%
\bibitem [{\citenamefont {Kaestner}\ \emph {et~al.}(2008)\citenamefont
  {Kaestner}, \citenamefont {Kashcheyevs}, \citenamefont {Hein}, \citenamefont
  {Pierz}, \citenamefont {Siegner},\ and\ \citenamefont
  {Schumacher}}]{Kaestner}%
  \BibitemOpen
  \bibfield  {author} {\bibinfo {author} {\bibfnamefont {B.}~\bibnamefont
  {Kaestner}}, \bibinfo {author} {\bibfnamefont {V.}~\bibnamefont
  {Kashcheyevs}}, \bibinfo {author} {\bibfnamefont {G.}~\bibnamefont {Hein}},
  \bibinfo {author} {\bibfnamefont {K.}~\bibnamefont {Pierz}}, \bibinfo
  {author} {\bibfnamefont {U.}~\bibnamefont {Siegner}}, \ and\ \bibinfo
  {author} {\bibfnamefont {H.~W.}\ \bibnamefont {Schumacher}},\ }\href@noop {}
  {\bibfield  {journal} {\bibinfo  {journal} {App. Phys. Lett. 92 192106}\ }
  (\bibinfo {year} {2008})}\BibitemShut {NoStop}%
\bibitem [{\citenamefont {Leicht}\ \emph {et~al.}(2011)\citenamefont {Leicht},
  \citenamefont {Mirovsky}, \citenamefont {Kaestner}, \citenamefont {Hohls},
  \citenamefont {Kashcheyevs}, \citenamefont {Kurganova}, \citenamefont
  {Zeitler}, \citenamefont {Weimann}, \citenamefont {Pierz},\ and\
  \citenamefont {Schumacher}}]{Leicht}%
  \BibitemOpen
  \bibfield  {author} {\bibinfo {author} {\bibfnamefont {C.}~\bibnamefont
  {Leicht}}, \bibinfo {author} {\bibfnamefont {P.}~\bibnamefont {Mirovsky}},
  \bibinfo {author} {\bibfnamefont {B.}~\bibnamefont {Kaestner}}, \bibinfo
  {author} {\bibfnamefont {F.}~\bibnamefont {Hohls}}, \bibinfo {author}
  {\bibfnamefont {V.}~\bibnamefont {Kashcheyevs}}, \bibinfo {author}
  {\bibfnamefont {E.~V.}\ \bibnamefont {Kurganova}}, \bibinfo {author}
  {\bibfnamefont {U.}~\bibnamefont {Zeitler}}, \bibinfo {author} {\bibfnamefont
  {T.}~\bibnamefont {Weimann}}, \bibinfo {author} {\bibfnamefont
  {K.}~\bibnamefont {Pierz}}, \ and\ \bibinfo {author} {\bibfnamefont {H.~W.}\
  \bibnamefont {Schumacher}},\ }\href@noop {} {\bibfield  {journal} {\bibinfo
  {journal} {Semicond. Sci. Tech. 26 055010}\ } (\bibinfo {year}
  {2011})}\BibitemShut {NoStop}%
\bibitem [{\citenamefont {Giblin}\ \emph {et~al.}(2012)\citenamefont {Giblin},
  \citenamefont {Kataoka}, \citenamefont {Fletcher}, \citenamefont {See},
  \citenamefont {Janssen}, \citenamefont {Griffiths}, \citenamefont {Jones},
  \citenamefont {Farrer},\ and\ \citenamefont {Ritchie}}]{Giblin}%
  \BibitemOpen
  \bibfield  {author} {\bibinfo {author} {\bibfnamefont {S.~P.}\ \bibnamefont
  {Giblin}}, \bibinfo {author} {\bibfnamefont {M.}~\bibnamefont {Kataoka}},
  \bibinfo {author} {\bibfnamefont {J.~D.}\ \bibnamefont {Fletcher}}, \bibinfo
  {author} {\bibfnamefont {P.}~\bibnamefont {See}}, \bibinfo {author}
  {\bibfnamefont {T.~J. B.~M.}\ \bibnamefont {Janssen}}, \bibinfo {author}
  {\bibfnamefont {J.~P.}\ \bibnamefont {Griffiths}}, \bibinfo {author}
  {\bibfnamefont {G.~A.~C.}\ \bibnamefont {Jones}}, \bibinfo {author}
  {\bibfnamefont {I.}~\bibnamefont {Farrer}}, \ and\ \bibinfo {author}
  {\bibfnamefont {D.~A.}\ \bibnamefont {Ritchie}},\ }\href@noop {} {\bibfield
  {journal} {\bibinfo  {journal} {Nature Comms. 3 1935}\ } (\bibinfo {year}
  {2012})}\BibitemShut {NoStop}%
\bibitem [{\citenamefont {Heiblum}\ \emph {et~al.}(1985)\citenamefont
  {Heiblum}, \citenamefont {Nathan}, \citenamefont {Thomas},\ and\
  \citenamefont {Knoedler}}]{Heiblum}%
  \BibitemOpen
  \bibfield  {author} {\bibinfo {author} {\bibfnamefont {M.}~\bibnamefont
  {Heiblum}}, \bibinfo {author} {\bibfnamefont {M.~I.}\ \bibnamefont {Nathan}},
  \bibinfo {author} {\bibfnamefont {D.~C.}\ \bibnamefont {Thomas}}, \ and\
  \bibinfo {author} {\bibfnamefont {C.~M.}\ \bibnamefont {Knoedler}},\
  }\href@noop {} {\bibfield  {journal} {\bibinfo  {journal} {Phys. Rev. Lett.
  55 2200}\ } (\bibinfo {year} {1985})}\BibitemShut {NoStop}%
\bibitem [{\citenamefont {\surname{Das Sarma}}\ and\ \citenamefont
  {Madhukar}(1980)}]{dasSarma}%
  \BibitemOpen
  \bibfield  {author} {\bibinfo {author} {\bibfnamefont {S.}~\bibnamefont
  {\surname{Das Sarma}}}\ and\ \bibinfo {author} {\bibfnamefont
  {A.}~\bibnamefont {Madhukar}},\ }\href@noop {} {\bibfield  {journal}
  {\bibinfo  {journal} {Phys. Rev. B 22 2823}\ } (\bibinfo {year}
  {1980})}\BibitemShut {NoStop}%
\bibitem [{\citenamefont {\surname{Das Sarma}}\ and\ \citenamefont
  {Campos}(1994)}]{dasSarma2}%
  \BibitemOpen
  \bibfield  {author} {\bibinfo {author} {\bibfnamefont {S.}~\bibnamefont
  {\surname{Das Sarma}}}\ and\ \bibinfo {author} {\bibfnamefont {V.~B.}\
  \bibnamefont {Campos}},\ }\href@noop {} {\bibfield  {journal} {\bibinfo
  {journal} {Phys. Rev. B 49, 1867}\ } (\bibinfo {year} {1994})}\BibitemShut
  {NoStop}%
\bibitem [{\citenamefont {Sivan}\ \emph {et~al.}(1989)\citenamefont {Sivan},
  \citenamefont {Heiblum},\ and\ \citenamefont {Umbach}}]{Sivan}%
  \BibitemOpen
  \bibfield  {author} {\bibinfo {author} {\bibfnamefont {U.}~\bibnamefont
  {Sivan}}, \bibinfo {author} {\bibfnamefont {M.}~\bibnamefont {Heiblum}}, \
  and\ \bibinfo {author} {\bibfnamefont {C.~P.}\ \bibnamefont {Umbach}},\
  }\href@noop {} {\bibfield  {journal} {\bibinfo  {journal} {Phys. Rev. Lett.
  63 9}\ } (\bibinfo {year} {1989})}\BibitemShut {NoStop}%
\bibitem [{\citenamefont {Taubert}\ \emph {et~al.}(2011)\citenamefont
  {Taubert}, \citenamefont {Tomaras}, \citenamefont {Schinner}, \citenamefont
  {Tranitz}, \citenamefont {Wegscheider}, \citenamefont {Kehrein},\ and\
  \citenamefont {Ludwig}}]{Taubert}%
  \BibitemOpen
  \bibfield  {author} {\bibinfo {author} {\bibfnamefont {D.}~\bibnamefont
  {Taubert}}, \bibinfo {author} {\bibfnamefont {C.}~\bibnamefont {Tomaras}},
  \bibinfo {author} {\bibfnamefont {G.~J.}\ \bibnamefont {Schinner}}, \bibinfo
  {author} {\bibfnamefont {H.~P.}\ \bibnamefont {Tranitz}}, \bibinfo {author}
  {\bibfnamefont {W.}~\bibnamefont {Wegscheider}}, \bibinfo {author}
  {\bibfnamefont {S.}~\bibnamefont {Kehrein}}, \ and\ \bibinfo {author}
  {\bibfnamefont {S.}~\bibnamefont {Ludwig}},\ }\href@noop {} {\bibfield
  {journal} {\bibinfo  {journal} {Phys. Rev. B 83, 235404}\ } (\bibinfo {year}
  {2011})}\BibitemShut {NoStop}%
\bibitem [{\citenamefont {Taubert}\ \emph {et~al.}(2010)\citenamefont
  {Taubert}, \citenamefont {Schinner}, \citenamefont {Tranitz}, \citenamefont
  {Wegscheider}, \citenamefont {Tomaras}, \citenamefont {Kehrein},\ and\
  \citenamefont {Ludwig}}]{Taubert2}%
  \BibitemOpen
  \bibfield  {author} {\bibinfo {author} {\bibfnamefont {D.}~\bibnamefont
  {Taubert}}, \bibinfo {author} {\bibfnamefont {G.~J.}\ \bibnamefont
  {Schinner}}, \bibinfo {author} {\bibfnamefont {H.~P.}\ \bibnamefont
  {Tranitz}}, \bibinfo {author} {\bibfnamefont {W.}~\bibnamefont
  {Wegscheider}}, \bibinfo {author} {\bibfnamefont {C.}~\bibnamefont
  {Tomaras}}, \bibinfo {author} {\bibfnamefont {S.}~\bibnamefont {Kehrein}}, \
  and\ \bibinfo {author} {\bibfnamefont {S.}~\bibnamefont {Ludwig}},\
  }\href@noop {} {\bibfield  {journal} {\bibinfo  {journal} {Phys. Rev. B 82
  161416R}\ } (\bibinfo {year} {2010})}\BibitemShut {NoStop}%
\bibitem [{\citenamefont {Bocquillon}\ \emph {et~al.}(2013)\citenamefont
  {Bocquillon}, \citenamefont {Freulon}, \citenamefont {Berroir}, \citenamefont
  {Degiovanni}, \citenamefont {Pla{\c{c}}ais}, \citenamefont {Cavanna},
  \citenamefont {Jin},\ and\ \citenamefont {F{\`{e}}ve}}]{Bocquillon}%
  \BibitemOpen
  \bibfield  {author} {\bibinfo {author} {\bibfnamefont {E.}~\bibnamefont
  {Bocquillon}}, \bibinfo {author} {\bibfnamefont {V.}~\bibnamefont {Freulon}},
  \bibinfo {author} {\bibfnamefont {J.~M.}\ \bibnamefont {Berroir}}, \bibinfo
  {author} {\bibfnamefont {P.}~\bibnamefont {Degiovanni}}, \bibinfo {author}
  {\bibfnamefont {B.}~\bibnamefont {Pla{\c{c}}ais}}, \bibinfo {author}
  {\bibfnamefont {A.}~\bibnamefont {Cavanna}}, \bibinfo {author} {\bibfnamefont
  {Y.}~\bibnamefont {Jin}}, \ and\ \bibinfo {author} {\bibfnamefont
  {G.}~\bibnamefont {F{\`{e}}ve}},\ }\href@noop {} {\bibfield  {journal}
  {\bibinfo  {journal} {Science 339 (6123) 1054}\ } (\bibinfo {year}
  {2013})}\BibitemShut {NoStop}%
\bibitem [{\citenamefont {Bennett}\ and\ \citenamefont
  {DiVincenzo}(2000)}]{Bennett}%
  \BibitemOpen
  \bibfield  {author} {\bibinfo {author} {\bibfnamefont {C.~H.}\ \bibnamefont
  {Bennett}}\ and\ \bibinfo {author} {\bibfnamefont {D.~P.}\ \bibnamefont
  {DiVincenzo}},\ }\href@noop {} {\bibfield  {journal} {\bibinfo  {journal}
  {Nature 404 247}\ } (\bibinfo {year} {2000})}\BibitemShut {NoStop}%
\bibitem [{\citenamefont {Fletcher}\ \emph {et~al.}(2013)\citenamefont
  {Fletcher}, \citenamefont {See}, \citenamefont {Howe}, \citenamefont
  {Pepper}, \citenamefont {Giblin}, \citenamefont {Griffiths}, \citenamefont
  {Jones}, \citenamefont {Farrer}, \citenamefont {Ritchie}, \citenamefont
  {Janssen},\ and\ \citenamefont {Kataoka}}]{Jon}%
  \BibitemOpen
  \bibfield  {author} {\bibinfo {author} {\bibfnamefont {J.~D.}\ \bibnamefont
  {Fletcher}}, \bibinfo {author} {\bibfnamefont {P.}~\bibnamefont {See}},
  \bibinfo {author} {\bibfnamefont {H.}~\bibnamefont {Howe}}, \bibinfo {author}
  {\bibfnamefont {M.}~\bibnamefont {Pepper}}, \bibinfo {author} {\bibfnamefont
  {S.~P.}\ \bibnamefont {Giblin}}, \bibinfo {author} {\bibfnamefont {J.~P.}\
  \bibnamefont {Griffiths}}, \bibinfo {author} {\bibfnamefont {G.~A.~C.}\
  \bibnamefont {Jones}}, \bibinfo {author} {\bibfnamefont {I.}~\bibnamefont
  {Farrer}}, \bibinfo {author} {\bibfnamefont {D.~A.}\ \bibnamefont {Ritchie}},
  \bibinfo {author} {\bibfnamefont {T.~J. B.~M.}\ \bibnamefont {Janssen}}, \
  and\ \bibinfo {author} {\bibfnamefont {M.}~\bibnamefont {Kataoka}},\
  }\href@noop {} {\bibfield  {journal} {\bibinfo  {journal} {Phys. Rev. Lett.
  111 216807}\ } (\bibinfo {year} {2013})}\BibitemShut {NoStop}%
\bibitem [{\citenamefont {Kataoka}\ \emph
  {et~al.}(2016{\natexlab{a}})\citenamefont {Kataoka}, \citenamefont {Johnson},
  \citenamefont {Emary}, \citenamefont {See}, \citenamefont {Griffiths},
  \citenamefont {Jones}, \citenamefont {Farrer}, \citenamefont {Ritchie},
  \citenamefont {Pepper},\ and\ \citenamefont {Janssen}}]{masaya}%
  \BibitemOpen
  \bibfield  {author} {\bibinfo {author} {\bibfnamefont {M.}~\bibnamefont
  {Kataoka}}, \bibinfo {author} {\bibfnamefont {N.}~\bibnamefont {Johnson}},
  \bibinfo {author} {\bibfnamefont {C.}~\bibnamefont {Emary}}, \bibinfo
  {author} {\bibfnamefont {P.}~\bibnamefont {See}}, \bibinfo {author}
  {\bibfnamefont {J.~P.}\ \bibnamefont {Griffiths}}, \bibinfo {author}
  {\bibfnamefont {G.~A.~C.}\ \bibnamefont {Jones}}, \bibinfo {author}
  {\bibfnamefont {I.}~\bibnamefont {Farrer}}, \bibinfo {author} {\bibfnamefont
  {D.~A.}\ \bibnamefont {Ritchie}}, \bibinfo {author} {\bibfnamefont
  {M.}~\bibnamefont {Pepper}}, \ and\ \bibinfo {author} {\bibfnamefont {T.~J.
  B.~M.}\ \bibnamefont {Janssen}},\ }\href@noop {} {\bibfield  {journal}
  {\bibinfo  {journal} {Phys. Rev. Lett. 116, 126803}\ } (\bibinfo {year}
  {2016}{\natexlab{a}})}\BibitemShut {NoStop}%
\bibitem [{\citenamefont {Emary}\ \emph {et~al.}(2016)\citenamefont {Emary},
  \citenamefont {Dyson}, \citenamefont {Ryu}, \citenamefont {Sim},\ and\
  \citenamefont {Kataoka}}]{Emary}%
  \BibitemOpen
  \bibfield  {author} {\bibinfo {author} {\bibfnamefont {C.}~\bibnamefont
  {Emary}}, \bibinfo {author} {\bibfnamefont {A.}~\bibnamefont {Dyson}},
  \bibinfo {author} {\bibfnamefont {S.}~\bibnamefont {Ryu}}, \bibinfo {author}
  {\bibfnamefont {H.-S.}\ \bibnamefont {Sim}}, \ and\ \bibinfo {author}
  {\bibfnamefont {M.}~\bibnamefont {Kataoka}},\ }\href@noop {} {\bibfield
  {journal} {\bibinfo  {journal} {Phys. Rev. B 93, 035436}\ } (\bibinfo {year}
  {2016})}\BibitemShut {NoStop}%
\bibitem [{\citenamefont {Komiyama}\ \emph {et~al.}(1992)\citenamefont
  {Komiyama}, \citenamefont {Hirai}, \citenamefont {Ohsawa}, \citenamefont
  {Matsuda}, \citenamefont {Sasa},\ and\ \citenamefont {Fujii}}]{Komiyama}%
  \BibitemOpen
  \bibfield  {author} {\bibinfo {author} {\bibfnamefont {S.}~\bibnamefont
  {Komiyama}}, \bibinfo {author} {\bibfnamefont {H.}~\bibnamefont {Hirai}},
  \bibinfo {author} {\bibfnamefont {M.}~\bibnamefont {Ohsawa}}, \bibinfo
  {author} {\bibfnamefont {Y.}~\bibnamefont {Matsuda}}, \bibinfo {author}
  {\bibfnamefont {S.}~\bibnamefont {Sasa}}, \ and\ \bibinfo {author}
  {\bibfnamefont {T.}~\bibnamefont {Fujii}},\ }\href@noop {} {\bibfield
  {journal} {\bibinfo  {journal} {Phys. Rev. B 45 11085}\ } (\bibinfo {year}
  {1992})}\BibitemShut {NoStop}%
\bibitem [{\citenamefont {Halperin}(1982)}]{Halperin}%
  \BibitemOpen
  \bibfield  {author} {\bibinfo {author} {\bibfnamefont {B.~I.}\ \bibnamefont
  {Halperin}},\ }\href@noop {} {\bibfield  {journal} {\bibinfo  {journal}
  {Phys. Rev. B 25 2185}\ } (\bibinfo {year} {1982})}\BibitemShut {NoStop}%
\bibitem [{Note1()}]{Note1}%
  \BibitemOpen
  \bibinfo {note} {This is the lithographic length, and we assume a $\pm 5\%$
  error because we do not know the exact path the electrons take. We note that
  the error in path length estimate does not contribute to the error in the
  measurements of phonon emission rate as it is cancelled out by both the
  velocity and length estimates. It does, however, contribute to the error in
  the estimate of the edge potential profile.}\BibitemShut {Stop}%
\bibitem [{\citenamefont {Waldie}\ \emph {et~al.}(2015)\citenamefont {Waldie},
  \citenamefont {See}, \citenamefont {Kashcheyevs}, \citenamefont {Griffiths},
  \citenamefont {Farrer}, \citenamefont {Jones}, \citenamefont {Ritchie},
  \citenamefont {Janssen},\ and\ \citenamefont {Kataoka}}]{joanna}%
  \BibitemOpen
  \bibfield  {author} {\bibinfo {author} {\bibfnamefont {J.}~\bibnamefont
  {Waldie}}, \bibinfo {author} {\bibfnamefont {P.}~\bibnamefont {See}},
  \bibinfo {author} {\bibfnamefont {V.}~\bibnamefont {Kashcheyevs}}, \bibinfo
  {author} {\bibfnamefont {J.~P.}\ \bibnamefont {Griffiths}}, \bibinfo {author}
  {\bibfnamefont {I.}~\bibnamefont {Farrer}}, \bibinfo {author} {\bibfnamefont
  {G.~A.~C.}\ \bibnamefont {Jones}}, \bibinfo {author} {\bibfnamefont {D.~A.}\
  \bibnamefont {Ritchie}}, \bibinfo {author} {\bibfnamefont {T.~J. B.~M.}\
  \bibnamefont {Janssen}}, \ and\ \bibinfo {author} {\bibfnamefont
  {M.}~\bibnamefont {Kataoka}},\ }\href@noop {} {\bibfield  {journal} {\bibinfo
   {journal} {Phys. Rev. B 92, 125305}\ } (\bibinfo {year} {2015})}\BibitemShut
  {NoStop}%
\bibitem [{\citenamefont {Johnson}\ \emph {et~al.}(2017)\citenamefont
  {Johnson}, \citenamefont {Fletcher}, \citenamefont {Humphreys}, \citenamefont
  {See}, \citenamefont {Griffiths}, \citenamefont {Jones}, \citenamefont
  {Farrer}, \citenamefont {Ritchie}, \citenamefont {Pepper}, \citenamefont
  {Janssen},\ and\ \citenamefont {Kataoka}}]{Johnson}%
  \BibitemOpen
  \bibfield  {author} {\bibinfo {author} {\bibfnamefont {N.}~\bibnamefont
  {Johnson}}, \bibinfo {author} {\bibfnamefont {J.~D.}\ \bibnamefont
  {Fletcher}}, \bibinfo {author} {\bibfnamefont {D.}~\bibnamefont {Humphreys}},
  \bibinfo {author} {\bibfnamefont {P.}~\bibnamefont {See}}, \bibinfo {author}
  {\bibfnamefont {J.}~\bibnamefont {Griffiths}}, \bibinfo {author}
  {\bibfnamefont {G.}~\bibnamefont {Jones}}, \bibinfo {author} {\bibfnamefont
  {I.}~\bibnamefont {Farrer}}, \bibinfo {author} {\bibfnamefont
  {D.}~\bibnamefont {Ritchie}}, \bibinfo {author} {\bibfnamefont
  {M.}~\bibnamefont {Pepper}}, \bibinfo {author} {\bibfnamefont
  {T.}~\bibnamefont {Janssen}}, \ and\ \bibinfo {author} {\bibfnamefont
  {M.}~\bibnamefont {Kataoka}},\ }\href@noop {} {\bibfield  {journal} {\bibinfo
   {journal} {Appl. Phys. Lett. 110 102105}\ } (\bibinfo {year}
  {2017})}\BibitemShut {NoStop}%
\bibitem [{\citenamefont {Kataoka}\ \emph
  {et~al.}(2016{\natexlab{b}})\citenamefont {Kataoka}, \citenamefont
  {Fletcher},\ and\ \citenamefont {Johnson}}]{Kataoka}%
  \BibitemOpen
  \bibfield  {author} {\bibinfo {author} {\bibfnamefont {M.}~\bibnamefont
  {Kataoka}}, \bibinfo {author} {\bibfnamefont {J.~D.}\ \bibnamefont
  {Fletcher}}, \ and\ \bibinfo {author} {\bibfnamefont {N.}~\bibnamefont
  {Johnson}},\ }\href {\doibase 10.1002/pssb.201600547} {\bibfield  {journal}
  {\bibinfo  {journal} {Phys. Status Solidii B 254 1521}\ } (\bibinfo {year}
  {2016}{\natexlab{b}}),\ 10.1002/pssb.201600547}\BibitemShut {NoStop}%
\bibitem [{Note2()}]{Note2}%
  \BibitemOpen
  \bibinfo {note} {In this example, $V_{G2} = -0.44$~V, $V_{G5} = -0.25$~V, $B
  = 11$~T, and $V_{G4} = -0.3$~V (short path, $S$) or $-0.65$~V (long path,
  $L$).}\BibitemShut {Stop}%
\bibitem [{Note3()}]{Note3}%
  \BibitemOpen
  \bibinfo {note} {We note that there is a smaller number of data points in
  Fig.~1(e) compared to Fig.~1(c) and (d). This is because we need both
  $1-P_{LO}^{l}$ and $v_{d}$ available in order to deduce $\Gamma _{LO}$ for a
  given experimental condition. The range in $E$ for which $1-P_{LO}^{l}$ can
  be measured is limited because, as the original electron emission is lowered
  below $E = -25$~meV (the energy of the electron that emits one phonon will be
  below -61~meV), we cannot reliably measure the height of the sub-step in
  $I_{d}$ as a spurious current due to a pick up of RF signal by the 2DEG
  starts to flow through the detector barrier when it is made too low. Also,
  $v_{d}$ measurement is difficult for less negative $V_{G5}$ and higher $E$,
  where the LO phonon emission rate is high, as there are not enough electrons
  reaching the detector with the original emission energy, so cannot be
  detected by the measurement of $I_{d}$.}\BibitemShut {Stop}%
\bibitem [{\citenamefont {Climente}\ \emph {et~al.}(2006)\citenamefont
  {Climente}, \citenamefont {Bertoni}, \citenamefont {Goldoni},\ and\
  \citenamefont {Molinari}}]{Climente}%
  \BibitemOpen
  \bibfield  {author} {\bibinfo {author} {\bibfnamefont {J.~I.}\ \bibnamefont
  {Climente}}, \bibinfo {author} {\bibfnamefont {A.}~\bibnamefont {Bertoni}},
  \bibinfo {author} {\bibfnamefont {G.}~\bibnamefont {Goldoni}}, \ and\
  \bibinfo {author} {\bibfnamefont {E.}~\bibnamefont {Molinari}},\ }\href@noop
  {} {\bibfield  {journal} {\bibinfo  {journal} {Phys. Rev. B 74 035313}\ }
  (\bibinfo {year} {2006})}\BibitemShut {NoStop}%
\end{thebibliography}%


\begin{thebibliography}{3}%
\makeatletter
\providecommand \@ifxundefined [1]{%
 \@ifx{#1\undefined}
}%
\providecommand \@ifnum [1]{%
 \ifnum #1\expandafter \@firstoftwo
 \else \expandafter \@secondoftwo
 \fi
}%
\providecommand \@ifx [1]{%
 \ifx #1\expandafter \@firstoftwo
 \else \expandafter \@secondoftwo
 \fi
}%
\providecommand \natexlab [1]{#1}%
\providecommand \enquote  [1]{``#1''}%
\providecommand \bibnamefont  [1]{#1}%
\providecommand \bibfnamefont [1]{#1}%
\providecommand \citenamefont [1]{#1}%
\providecommand \href@noop [0]{\@secondoftwo}%
\providecommand \href [0]{\begingroup \@sanitize@url \@href}%
\providecommand \@href[1]{\@@startlink{#1}\@@href}%
\providecommand \@@href[1]{\endgroup#1\@@endlink}%
\providecommand \@sanitize@url [0]{\catcode `\\12\catcode `\$12\catcode
  `\&12\catcode `\#12\catcode `\^12\catcode `\_12\catcode `\%12\relax}%
\providecommand \@@startlink[1]{}%
\providecommand \@@endlink[0]{}%
\providecommand \url  [0]{\begingroup\@sanitize@url \@url }%
\providecommand \@url [1]{\endgroup\@href {#1}{\urlprefix }}%
\providecommand \urlprefix  [0]{URL }%
\providecommand \Eprint [0]{\href }%
\providecommand \doibase [0]{http://dx.doi.org/}%
\providecommand \selectlanguage [0]{\@gobble}%
\providecommand \bibinfo  [0]{\@secondoftwo}%
\providecommand \bibfield  [0]{\@secondoftwo}%
\providecommand \translation [1]{[#1]}%
\providecommand \BibitemOpen [0]{}%
\providecommand \bibitemStop [0]{}%
\providecommand \bibitemNoStop [0]{.\EOS\space}%
\providecommand \EOS [0]{\spacefactor3000\relax}%
\providecommand \BibitemShut  [1]{\csname bibitem#1\endcsname}%
\let\auto@bib@innerbib\@empty
\bibitem [{\citenamefont {Kataoka}\ \emph
  {et~al.}(2016{\natexlab{a}})\citenamefont {Kataoka}, \citenamefont {Johnson},
  \citenamefont {Emary}, \citenamefont {See}, \citenamefont {Griffiths},
  \citenamefont {Jones}, \citenamefont {Farrer}, \citenamefont {Ritchie},
  \citenamefont {Pepper},\ and\ \citenamefont {Janssen}}]{masaya}%
  \BibitemOpen
  \bibfield  {author} {\bibinfo {author} {\bibfnamefont {M.}~\bibnamefont
  {Kataoka}}, \bibinfo {author} {\bibfnamefont {N.}~\bibnamefont {Johnson}},
  \bibinfo {author} {\bibfnamefont {C.}~\bibnamefont {Emary}}, \bibinfo
  {author} {\bibfnamefont {P.}~\bibnamefont {See}}, \bibinfo {author}
  {\bibfnamefont {J.~P.}\ \bibnamefont {Griffiths}}, \bibinfo {author}
  {\bibfnamefont {G.~A.~C.}\ \bibnamefont {Jones}}, \bibinfo {author}
  {\bibfnamefont {I.}~\bibnamefont {Farrer}}, \bibinfo {author} {\bibfnamefont
  {D.~A.}\ \bibnamefont {Ritchie}}, \bibinfo {author} {\bibfnamefont
  {M.}~\bibnamefont {Pepper}}, \ and\ \bibinfo {author} {\bibfnamefont {T.~J.
  B.~M.}\ \bibnamefont {Janssen}},\ }\href@noop {} {\bibfield  {journal}
  {\bibinfo  {journal} {Phys. Rev. Lett. 116, 126803}\ } (\bibinfo {year}
  {2016}{\natexlab{a}})}\BibitemShut {NoStop}%
\bibitem [{\citenamefont {Emary}\ \emph {et~al.}(2016)\citenamefont {Emary},
  \citenamefont {Dyson}, \citenamefont {Ryu}, \citenamefont {Sim},\ and\
  \citenamefont {Kataoka}}]{Emary}%
  \BibitemOpen
  \bibfield  {author} {\bibinfo {author} {\bibfnamefont {C.}~\bibnamefont
  {Emary}}, \bibinfo {author} {\bibfnamefont {A.}~\bibnamefont {Dyson}},
  \bibinfo {author} {\bibfnamefont {S.}~\bibnamefont {Ryu}}, \bibinfo {author}
  {\bibfnamefont {H.-S.}\ \bibnamefont {Sim}}, \ and\ \bibinfo {author}
  {\bibfnamefont {M.}~\bibnamefont {Kataoka}},\ }\href@noop {} {\bibfield
  {journal} {\bibinfo  {journal} {Phys. Rev. B 93, 035436}\ } (\bibinfo {year}
  {2016})}\BibitemShut {NoStop}%
\bibitem [{\citenamefont {Kataoka}\ \emph
  {et~al.}(2016{\natexlab{b}})\citenamefont {Kataoka}, \citenamefont
  {Fletcher},\ and\ \citenamefont {Johnson}}]{Kataoka}%
  \BibitemOpen
  \bibfield  {author} {\bibinfo {author} {\bibfnamefont {M.}~\bibnamefont
  {Kataoka}}, \bibinfo {author} {\bibfnamefont {J.~D.}\ \bibnamefont
  {Fletcher}}, \ and\ \bibinfo {author} {\bibfnamefont {N.}~\bibnamefont
  {Johnson}},\ }\href {\doibase 10.1002/pssb.201600547} {\bibfield  {journal}
  {\bibinfo  {journal} {Phys. Status Solidii B 254 1521}\ } (\bibinfo {year}
  {2016}{\natexlab{b}}),\ 10.1002/pssb.201600547}\BibitemShut {NoStop}%
\end{thebibliography}%
	
\end{document}


\title{Supplemental Material to LO-phonon emission rate of hot electrons from an on-demand single-electron source in a GaAs/AlGaAs heterostructure}

\author  {N. Johnson}
\email[]{nathan.johnson@lab.ntt.co.jp}
\altaffiliation[Present address: ]{NTT Basic Research Laboratories, NTT Corporation, 3-1 Morinosato Wakamiya, Atsugi, Kanagawa 243-0198, Japan}
\affiliation{National Physical Laboratory, Hampton Road, Teddington, Middlesex TW11 0LW, United Kingdom}
\affiliation{London Centre for Nanotechnology, and Department of Electronic and Electrical Engineering, University College London, Torrington Place, London, WC1E 7JE, United Kingdom}

\author {C. Emary}
\affiliation{Joint Quantum Centre Durham-Newcastle, School of Mathematics and Statistics, Newcastle University, Newcastle upon Tyne NE1 7RU, United Kingdom.}
\author{S. Ryu}
\affiliation{Department of Physics, Korea Advanced Institute of Science and Technology, Daejeon 305-701, Republic of Korea}
\author{H.-S. Sim}
\affiliation{Department of Physics, Korea Advanced Institute of Science and Technology, Daejeon 305-701, Republic of Korea}
\author {P. See}
\affiliation{National Physical Laboratory, Hampton Road, Teddington, Middlesex TW11 0LW, United Kingdom}
\author {J.D. Fletcher}
\affiliation{National Physical Laboratory, Hampton Road, Teddington, Middlesex TW11 0LW, United Kingdom}
\author{J.P. Griffiths}
\affiliation {Cavendish Laboratory, University of Cambridge, J.J. Thomson Avenue, Cambridge CB3 0HE, United Kingdom}
\author{G.A.C. Jones}
\affiliation{Cavendish Laboratory, University of Cambridge, J.J. Thomson Avenue, Cambridge CB3 0HE, United Kingdom}
\author{I. Farrer}
\affiliation{Department of Electronic and Electrical Engineering, University of Sheffield, Mappin Street, Sheffield S1 3JD, United Kingdom}
\author{D.A. Ritchie}
\affiliation{Cavendish Laboratory, University of Cambridge, J.J. Thomson Avenue, Cambridge CB3 0HE, United Kingdom}
\author{M. Pepper}
\affiliation{London Centre for Nanotechnology, and Department of Electronic and Electrical Engineering, University College London, Torrington Place, London, WC1E 7JE, United Kingdom}
\author{T.J.B.M Janssen}
\affiliation{National Physical Laboratory, Hampton Road, Teddington, Middlesex TW11 0LW, United Kingdom}
\author{M. Kataoka}
\email[]{masaya.kataoka@npl.co.uk}
\affiliation{National Physical Laboratory, Hampton Road, Teddington, Middlesex TW11 0LW, United Kingdom}

\begin{abstract}
\end{abstract}

\maketitle
\section{Determination of path length $l$}

In the analysis we calculated the LO emission rate as $\Gamma_{LO} =  - \frac{v_{d}}{l} \rm{ln}$$(1-P_{LO}^{l})$ using the path length $l = 20 \mu$m, which is the length of the loop alone.
As stated in the paper, there is no contribution to the rate from emission in the length along the gate $\rm{G_{4}}$.
We justify this statement here, first showing our measurement of $v_{d}$ is independent of this path length, and then the survival $1-P_{LO}^{l}$ also. 

Fig.~S1 plots the time of arrival of electrons at the detector $\rm{G_{3}}$ as a function of the deflection gate voltage $V_{G4}$.
The interval marked shows the time of flight $\tau_{2}$, with which the drift velocity $v_{d}$ is calculated, and this is understood to be the time of flight around the loop length $l$ alone.
The argument follows that of Ref.~\cite{masaya}, and Fig.~S1 is equivalent to Fig.~2(f) of Ref.~\cite{masaya}.
We divide the long path into three sections, with three different average drift velocities $v_{1,2,3}$ (see image in Fig.~S1). 
The first is the length $L_{1}$, alongside $\rm{G_{4}}$, and the second is the length $L_{2} = l$, the loop length.
The third is the section in common with the short path, which is not of importance as the time of flight in this section is cancelled out in our measurements.
As the path alongside the gate $\rm{G_{4}}$ has a different potential profile to that in the loop, we identify two contributions to the time of flight, one due to the profile of the loop and one due to the profile along the gate $\rm{G_{4}}$, and so we write the time of flight $\tau_{2}$ as $\tau = \tau_{1}+\tau_{2}=\frac{2L_{1}}{v_{1}} + \frac{L_{2}}{v_{2}}$ as marked on Fig.~S1.
At the most negative $V_{G4}$, the potential profile along $\rm{G_{4}}$ should be very steep, resulting in a high $v_{1}$ and so negligible $\tau_{1}$. 
In these cases, we expect the profile to be much sharper than the measured profiles (main paper, Fig.~2(b)), 
and $v_{d}$ (main paper, Fig.~1(d)) is dominated by the velocity in the loop.
For this work, we used $V_{G4} = -0.3$ or $-0.65$~V, for all measurements, which is within the saturated arrival times for all energies.

\begin{figure}[h!]
	\centering
	\includegraphics[scale=0.3]{figs1}
	\caption{Determination of the time of flight $\tau$, with which we calculate the drift velocity $v_{d}$.}	
\end{figure}

\begin{figure}[h!]
	\centering
	\includegraphics[scale=0.3]{figs2}
	\caption{Survival fraction $1-P_{0}^{L}$, as a function of energy, for two $V_{G4}$ values, $-0.5$~V and $-0.65$~V. 
		Overlying traces show no substantial change in LO-phonon emission with $V_{G4}$, implying that the path length along $\rm{G_{4}}$ does not contribute to the measured LO-phonon emission rate $\Gamma_{LO}$.}	
\end{figure}

Next, we show that the phonon emission is also negligible in the region along $\rm{G_{4}}$ due to the short time of flight, which is essential in this work if the terms in the rate $\Gamma_{LO}$ are to be measured using the same path length.
We use the same argument for the survival fraction $1-P_{LO}^{L}$ as we did for $v_{d}$. 
We measure the survival fraction (as plotted in the main paper, Fig.~1(c)) using a value of $V_{G4}$ such that the time of arrival does not change with $V_{G4}$.
Fig.~S2 shows the survival fraction in the long path $1-P_{0}^{L}$ as a function of energy for two values of $V_{G5}$, $-0.5$~V,
and $-0.65$~V, which is the value taken in measurement of $1-P_{LO}^{L}$ (and $v_{d}$).
(This data is taken at $V_{G5} = -0.25$~V and $B = 11$~T.)
We see no difference between the traces, implying there is no phonon emission observed in the path along $\rm{G_{4}}$.
This can be expected because the velocity in this region is very high (the time of flight is very small), and so the chance of an LO-phonon emission is minimal.
Hence we can conclude that the loop length alone of $20~\mu$m is the length in which we measure $\Gamma_{LO}$ through measurement of $v_{d}$ and $1- P_{LO}^{l}$.

\section{Derivation for the 1 LO-phonon emission arrival time distribution $A_{LO}(t)$}

In the main paper, Fig. 2(a), the blue trace plots the expected form of the arrival time distribution $A_{LO}(t)$ for electrons emitting a single LO-phonon, with emission uniform throughout the loop length $l$.
We now derive the form of this curve, following the method presented in Ref.~\cite{Emary}.

We begin with the ``coupled Boltzmann-like" equations, eqn.~16 of Ref.~\cite{Emary}, which are:

\begin{equation}
\frac{\partial}{\partial t}\varrho_{n}^{(m)} + v_{n}^{(m)} \frac{\partial}{\partial x}\varrho_{n}^{(m)} = -\left(1-\delta_{m,M}\right)\Gamma_{n}^{(m)}\varrho_{n}^{(m)} + \left(1-\delta_{m,0}\right) \sum_{n'}\Gamma_{nn'}^{(m-1)}\varrho_{n'}^{m-1}
\label{eqn:boltzmann}
\end{equation}
\noindent
where $\varrho(x,t)$ is the classical probability distribution of finding an electron, having emitted $m$ phonons from edge state (Landau Level) $n$ at position $x$ and time $t$, $v_{n}^{(m)}$ the electron velocity in the subband $n$ with energy $E^{(m)}$.
$M$ is the maximum number of LO-phonons that can be emitted before the electron energy falls to less than $\hbar \omega_{LO}$ from the subband bottom, and $\Gamma_{n}^{(m)}$ is the $m$th LO-phonon emission rate from subband $n$.

In our case, we consider only a single emission from the outermost edge channel, so $M = 1$ and $n = 0$, and we simplify Eqn.~\ref{eqn:boltzmann} as

\begin{equation}
\begin{split}
\frac{\partial }  {\partial t} \varrho^{(0)} + v^{(0)} \frac{\partial}{\partial x} \varrho^{(0)} = -\Gamma \varrho^{(0)};\\
\frac{\partial}{\partial t} \varrho^{(1)} + v^{(1)} \frac{\partial}{\partial x} \varrho^{(1)} = \Gamma \varrho^{(0)}.
\end{split}
\end{equation}
\noindent
with initial Gaussian distributions \cite{Emary, Kataoka}

\begin{equation}
\begin{split}
\varrho^{(0)}(x,0) &= f_{0}(x) = \frac{1}{\sqrt{2 \pi}w} \mathrm{exp}\left(-\frac{1}{2} \left(\frac{x}{w}\right)^{2}\right);\\
\varrho^{(1)}(x,0) &= 0
\end{split}
\end{equation}
\noindent
where $w$ is the standard deviation.

These equations solve to
\begin{equation*}
\varrho^{(0)}(x,t) = e^{-\Gamma t}f_{0}\left(x - v^{(0)}t\right);
\end{equation*}
\begin{equation}
\centering
\begin{split}
\varrho^{(1)}(x,t) = \frac{\Gamma}{2\left( v^{(0)}-v^{(1)} \right)}\mathrm{exp} \left( \frac{\Gamma}{2\left( v^{(0)} - v^{(1)}\right)^{2}} \left[ w^{2}\Gamma - 2\left(v^{(0)} - v^{(1)}\right)\left( x - v^{(1)}t\right) \right] \right) \\
\times \left( \mathrm{Erf}\left[ 
	\frac{w^{2}\Gamma -  \left(v^{(0)} - v^{(1)}\right) \left(x - v^{(0)}t\right)}{\sqrt{2}w \left( v^{(0)}-v^{(1)}\right)}\right] - 
\mathrm{Erf}\left[ \frac{w^{2}\Gamma -  \left(v^{(0)} - v^{(1)}\right) 
\left(x - v^{(1)}t\right)}
{\sqrt{2}w 
	\left( v^{(0)}-v^{(1)}\right)}\right] \right)
\end{split}
\end{equation}
\noindent
with the arrival time distribution $A_{LO}(x,t)$ given by

\begin{equation}
\begin{split}
A^{(0)}(x,t) = v^{(0)} \varrho^{(0)}(x,t);\\
A^{(1)}(x,t) = v^{(1)} \varrho^{(1)}(x,t);
\end{split}
\end{equation}

Finally, we evaluate $A_{LO}(x,t)$ at the position of the detector barrier $x_{d} = 28\mu$m so we plot $A_{LO}(t) = A^{(1)}(x=x_{d}, t)$ on Fig.~2(a).
In Fig.~2(a), $A_{LO}(t)$ is calculated using parameters 
$v^{(0)} = 11.6 \times 10^{4}$~m/s,
$v^{(1)} = 8.7 \times 10^{4}$~m/s,
$\Gamma = 6.3 \times 10^{9}$~/s,
$w = v^{(0)} \times 1$~ps $\approx 10$~ps.

\bibliography{phonons_reduced}